\documentstyle[11pt,paspconf,psfig]{article}

\input epsf.sty 
\begin{document}

\title{On the Blue Tail of Horizontal Branch Stars}

\author{G. Bono\altaffilmark{1}}
\affil{Osservatorio Astronomico di Roma, Via Frascati 33, 00040 Monteporzio,
Italy, {\tt bono@coma.mporzio.astro.it}}
\author{S. Cassisi}
\affil{Osservatorio Astronomico di Collurania, Via M. Maggini, 64100 Teramo,
Italy, {\tt cassisi@astrte.te.astro.it}}

\altaffiltext{1}{On leave from Osservatorio Astronomico di Trieste, Via G.B. 
Tiepolo 11, 34131 Trieste, Italy}

\begin{abstract}
We discuss theoretical predictions for ZAHB models constructed either by 
including or neglecting the He extra-mixing effect recently suggested by 
Sweigart (1997). The comparison in the $log T_e -log g$ plane suggests 
that within current observational uncertainties canonical hot HB models 
based on new input physics agree with empirical data in Galactic globular 
clusters and in the field.  
We also briefly discuss the impact of the new class of variable stars
EC14026 on constraining the parameters of blue HB stars and their 
distribution along the tail.  
\end{abstract}

% Keywords should be included, but they are not printed in the hardcopy.
% An attempt to select keywords following the U of Chicago Press subject
% headings (available on-line at http://www.noao.edu/apj/keywords96.html)
% would be greatly appreciated!
\keywords{Galaxy: halo --- ISM: clouds}

%%%%%%%%%%%%%%%%%%%%%%%%%%%%%%%%%%%%%%%%%%%%%%%%%%%%%%%%%%%%%%%%%%%%%%%%%%
\section{Introduction}

Only with a n\"aive approach to stellar astrophysics one could not realize 
that Horizontal Branch (HB) stars are the crossroad of several astrophysical
problems concerning the evolutionary properties and the final fate of 
low-mass stars as well as a fundamental laboratory for estimating 
important astrophysical parameters such as primordial He content 
by means of the $R$ parameter, stellar distances via RR Lyrae luminosity 
and ages of Globular Clusters (GCs) by means of the $\Delta{V}$ method 
(see for a comprehensive review Castellani 1999). 
Moreover, blue HB stars are currently adopted for tracing the luminous 
mass distribution of the Galactic halo, and in turn for constraining 
the primordial structure of the Galaxy (Kinman et al. 1996; 
Sluis \& Arnold 1998). 

The physical mechanisms governing core He burning evolutionary 
phases in low-mass stars were established in the seventies by 
Castellani et al. (1969), Iben \& Rood (1970), 
and by Sweigart \& Gross (1978). After these pioneering papers 
the evolutionary scenario has been soundly supplemented by several 
thorough investigations aimed at improving the input physics such as 
radiative opacities, equation of state and nuclear reaction rates  
(Dorman et al. 1993; Castellani et al. 1991; Cassisi et al. 1998). 
Thanks to large grids of HB models it has also been possible to account 
for the dependence of HB and post-HB evolutionary phases on chemical 
composition and therefore to develop a homogeneous theoretical 
framework for He burning phases (Dorman et al. 1993; 
Cassisi et al. 1998). During the last few years the HB scenario has 
undergone some sudden jerks due to new observational and theoretical 
results. 

$\bullet$ New HST photometric data collected by Rich et al. (1997) 
disclosed an interesting feature of blue HB stars in Galactic GCs. 
Thanks to the accuracy of these data they suggested that 
the luminosity of HB stars in NGC6388 and NGC6441 is tilted, 
i.e. when moving from low to high temperature the HB luminosity 
undergoes an increase of the order of $\Delta{V}=0.5$ mag.  
This observational evidence was independently supported by Grundahl
et al. (1998) who found a similar feature in M13 HB stars by adopting 
Str\"omgren photometric data. 

$\bullet$ Further observational evidence recently brought out is the 
occurrence of hot HB stars in some metal-rich GCs such as 47 Tuc and 
NGC362 (O'Connell et al. 1997; Dorman et al. 1997). This feature is at odds 
with predictions provided by canonical HB models, since it has been generally 
assumed so far that the blue tail is the finger-print of metal-poor clusters. 

$\bullet$ New photometric data also suggest that the gaps which appear  
along the blue tail of some GCs such as NGC2808, M80 and M13 are real 
(Sosin et al. 1997; Ferraro et al. 1998), thus confirming the results 
obtained by Newell (1973) for hot HB stars in the field. This evidence has 
been soundly supplemented by the recent discovery of a new class of variable 
stars called EC14026 by Kilkenny et al. (1997), which are characterized 
by surface gravities and effective temperatures typical of 
Extreme HB (EHB) stars. 
 
The new data stimulated several theoretical investigations which accounted 
for new physical mechanisms in explaining the HB morphology and the 
evolutionary properties of EHB stars. 
In particular, Sweigart (1997, 1998) suggested that the chemical 
anomalies observed in globular cluster red giants can be due to an 
extra-mixing which takes place inside their envelope. 
If this mechanism, which could be caused by internal rotation, 
is actually at work in real stars, the He abundance in the stellar 
envelope should also be enhanced. The increase in the He content 
brings about that these noncanonical HB models are, at fixed stellar 
mass, both brighter and bluer than the canonical ones. 
This effect is not constant along the HB since fast rotators undergo  
larger He enhancement than slow rotators. This gradient would imply 
that red HB stars are marginally affected by this phenomenon (slow 
rotators), whereas blue HB stars are strongly affected by extra-mixing 
(fast rotators). 
The main outcomes of this scenario have been exhaustively investigated 
by Sweigart (1998) and can be summarized as follows: 
1) the extra-mixing hypothesis accounts for tilted HB morphology;  
2) in comparison with canonical models the noncanonical ones predict, 
at fixed ZAHB effective temperature, a larger envelope mass and 
therefore overcome the fine tuning of mass loss efficiency necessary 
for producing EHB stars.  

%%%%%%%%%%%%%%%%%%%%%%%%%%%%%%%%%%%%%%%%%%%%%%%%%%%%%%%%%%%%%%%%%%%%%%%%%%
\section{Discussion}

For comparing in detail theory with observations,  
Figure 1 shows in the $log T_e - log g$ plane a comparison between HB 
models and current available data for hot HB stars in GCs. For avoiding 
any misleading effect due to the chemical composition the sample was 
selected by taking into account data of clusters for which accurate 
spectroscopic determinations of both Fe and $\alpha$-elements were 
available. The long dashed line in the upper panel for [M/H]=-1.3 
refer to HB models {\em \`a la} Sweigart, i.e. models which account 
for a He extra-mixing of the order of $\Delta X=0.10$. Solid and dashed 
lines show the ZAHB and the central He exhaustion of HB models which 
include new input physics and neglect He extra-mixing 
(see Bono et al. 1999). Data plotted in this figure 
suggest that canonical HB models are, within observational errors, 
in fair agreement with empirical data. At the same time, it is 
worth noting that gravities predicted by HB models with  
He extra-mixing are somewhat smaller than the observed gravities 
in NGC6752, M5, and M3. On the other hand, gravities of canonical 
models are slightly 
larger than the empirical ones in the more metal-rich cluster of our 
sample (NGC288). The main outcome of this comparison is that 
spectroscopic data marginally support the extra-mixing scenario.  

\begin{figure}
\hbox{
\centerline{\psfig{file=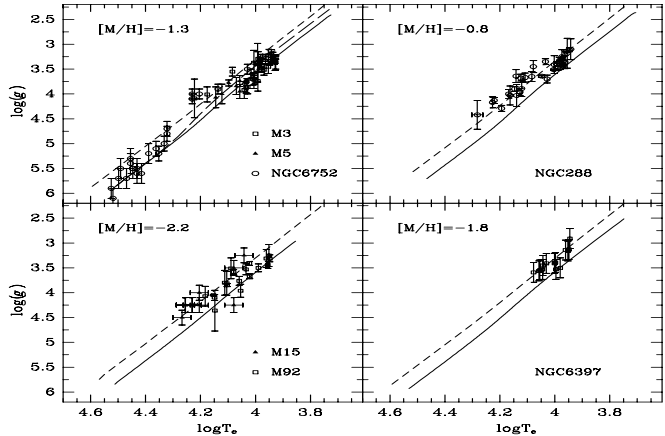,height=6cm,width=9.0cm}}
}
\caption{Comparison in the $log T_e - log g$ plane between theoretical 
HB models (solid line: ZAHB, dashed line: central He exhaustion) and 
empirical data. Spectroscopic measurements were collected by Crocker 
et al. (1988) for M3, M5, M15, M92 NGC288; de Boer et al. (1995) for 
NGC6397 and Moehler et al. (1997) for NGC6752. The long-dashed line 
in the top left panel refers to ZAHB models which account for He 
extra-mixing.}
\end{figure}

As a further test of the canonical HB evolutionary scenario, Figure 2 
shows in the $T_e - log g$ plane the comparison with field EHB stars 
collected by Saffer et al. (1994). 
Data plotted in this figure suggest that within current observational 
uncertainties -we adopted as conservative estimates $\sigma(g)=\pm 0.2$ dex 
and $\sigma(T_e)=\pm 1000$ K- theory and observations are in reasonable
agreement. In fact, a large number of these objects is located between 
the He zero age main sequence and the two ZAHBs for $Y_{MS}=0.23$, Z=0.0003 and 
for $Y_{MS}=0.28$, Z=0.02. The dashed line shows the loci along which HB 
structures of the former composition exhaust central He burning. 
The three evolutionary tracks for M=0.508, 0.515 and 0.520 $M_\odot$ suggest 
that even for post-HB evolutionary phases theoretical predictions 
agree with observational data. 

\begin{figure}
\hbox{
\centerline{\psfig{file=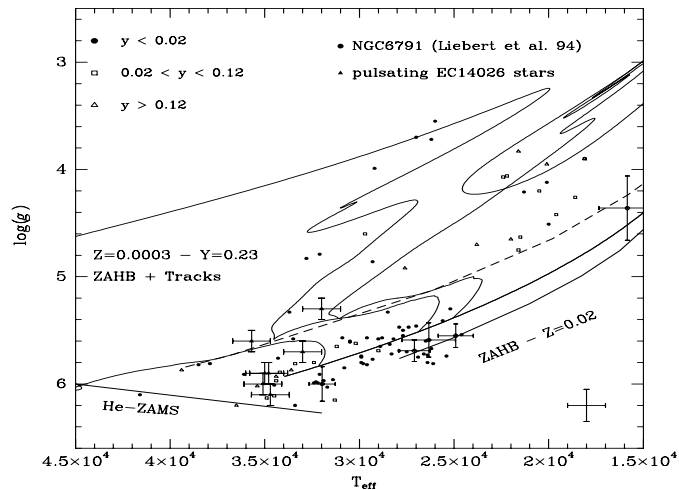,height=7cm,width=9.0cm}}
}
\caption{Comparison in the $T_e - log g$ plane between theoretical 
HB models and field EHB stars collected by Saffer et al. (1994). The 
data (full triangles) for pulsating EC 14026 stars as collected by 
O'Donoghue et al. (1998) have been also plotted. See text for further 
details.}
\end{figure}

An interesting feature of EHB stars recently discovered by Kilkenny et 
al. (1997) and by O'Donoghue et al. (1998 and references therein) is that 
some of these objects are variable stars. 
In Figure 2 we also plotted the variables that have 
been identified up to now. These objects can play a fundamental role
in understanding the evolutionary properties of EHB stars. In fact, 
the comparison between pulsation theory and observations can supply an 
independent estimate of astrophysical parameters governing the pulsation 
behavior of these variables. Moreover, the pulsation 
destabilization could also explain the occurrence of gap(s) along 
the blue tail. The EHB stars present a very thin envelope and therefore 
even small perturbations of the outermost layers can cause an increase in 
the efficiency of mass loss. The plausible consequence of this effect 
is that the gap(s) might be {\em region(s) of avoidance} for EHB stars. 
Unfortunately, we still 
lack a firm explanation of the physical mechanisms which drive their   
pulsation instability. On the basis of a detailed analysis of light 
curves and of linear pulsation models Stobie et al. (1997) suggested 
that both radial and nonradial modes should excited but the models 
they computed are pulsationally stable. 

In order to test whether the extra-mixing scenario suggested by 
Sweigart account for the pulsation behavior and modal stability 
of these variables we constructed several sequences of linear, nonadiabatic 
models by adopting a wide range of effective temperatures, luminosities, 
stellar masses and chemical compositions ranging from Y=0.24, Z=0.0001
to Y=0.34, Z=0.02. 
However, as already found by Stobie et al. (1997), the models are 
pulsationally stable. These numerical experiments suggest that the He 
abundance marginally affects the pulsation instability of EC14026 stars,
and in turn that the extra-mixing scenario does not help to explain 
their behavior.

Finally, we mention that a set of models at solar composition constructed 
by artificially enhancing (by 50\%) the opacity bump located close to  
$2.5\times10^5$ K are pulsationally unstable for temperatures ranging from
29000 to 26000 K. A different {\em ad hoc} mechanism for destabilizing 
these stars was suggested by Charpinet et al. (1996) but we still lack 
a straightforward understanding of the intimate nature of EC14026 stars.  
Certainly photometric and spectroscopic surveys of the Galactic halo 
currently undertaken can supply more tight constraints on the region 
of the HR diagram in which EHB stars are pulsationally unstable.


\begin{references}
\reference Bono, G., Cassisi, S. \& Castellani, V. 1999, {\sl in preparation}
\reference Cassisi, S., Castellani, V., Degl'Innocenti, S. \& Weiss, A. 1998, \aaps, 129, 267
\reference Castellani, V. 1999, in Globular Clusters,  (Cambridge: Cambridge 
University Press), in press  
\reference Castellani, V., Chieffi, A. \& Pulone, L. 1991, \apjs, 76, 911 
\reference Castellani, V., Giannone, P., Renzini, A. 1969, \apss, 3, 518
\reference Charpinet, S., Fontaine, G., Brassard, P. \& Dorman, B. 1996, \apj, 471, L103
\reference Crocker, D.A., Rood, R.T. \& O'Connell, R.W. 1988, \apj, 332, 236
\reference de Boer, K.S., Schmidt, J.H.K. \& Heber, U. 1995, \aap, 303, 95
\reference Dorman, B., Rood, R.T. \& O'Connell, R.W. 1993, \apj, 419, 596 
\reference Dorman, B. et al. 1997, \apjl, 480, 31
\reference Ferraro, F.R., Paltrinieri, B., Fusi Pecci, F., Rood, R.T. \& Dorman, B. 1998, \apj, 500, 311
\reference Grundahl, F., Vandenberg, D.A. \& Andersen, M.I. 1998, \apjl, 500, 179 
\reference Iben, I.Jr. \& Rood, R.T. 1970, \apj, 161, 587
\reference Kilkenny, D., Koen, C., O'Donoghue, D. \& Stobie, R.S. 1997, \mnras, 285, 640
\reference Kinman, T.D. et al.  1996, \aj, 111, 1164 
\reference Liebert J., Saffer R.A. \& Green E.M. 1994, AJ, 107, 1408
\reference Moehler, S., Heber, U. \& Rupprecht, G. 1997, \aap, 319, 109
\reference Newell, E.B. 1973, \apjs, 26, 37
\reference O'Connell, R.W. et al. 1997, \aj, 114, 1982
\reference O'Donoghue, D., Koen, C., Kilkenny, D., Stobie, R.S., Lynas-Gray, A.E. \& Kawaler, S.D. 1998, Balt.A., 7, 313
\reference Rich, R.M. et al. 1997, \apjl, 484, 25 
\reference Saffer, R.A., Bergeron, P., Koester, D. \& Liebert, J. 1994, \apj, 432, 351
\reference Sluis, A.P.N. \& Arnold, R.A. 1998, \mnras, 297, 732
\reference Sosin, C. et al. 1997, \apjl, 480, 35
\reference Stobie, R.S., Kawaler, S.D., Kilkenny, D., O'Donoghue, D. \& Koen, C. 1997,
\mnras, 285, 651
\reference Sweigart, A. V. 1997, \apj, 474, L23
\reference Sweigart, A. V. 1998, in The Third Conference on Faint Blue Stars, 
ed. A.G.D. Philip, J. Liebert \& R.A. Saffer (Cambridge: Cambridge Univ. 
Press), in press
\reference Sweigart, A. V. \& Gross, P.G. 1978, \apjs, 36, 405
%\reference Sweigart, A.V. \& Catelan, M. 1998, ASP Conf. Ser. 135, in "A half century of stellar pulsation interpretation: A tribute to A.N. Cox", eds. P.A. Bradley \& J.A. Guzik, (San Francisco:ASP), 39 
\end{references}
\end{document}